\documentclass[aps,twocolumn,pra,superscriptaddress,tightenlines]{revtex4}
\usepackage{mathrsfs}
\usepackage{amsmath}
\usepackage{graphicx}
\usepackage{epsfig}
\usepackage{txfonts}
\usepackage{subfigure}
\usepackage{amsfonts}
\usepackage{CJK}
\usepackage{color}
\usepackage[colorlinks,citecolor=blue]{hyperref}
\usepackage{bm}
\begin{document}

\begin{CJK*}{GBK}{song}
\title{Antisymmetry breaking coupling-enhanced sensing of quantum reservoirs }
\author{Ji-Bing Yuan\footnote{ Email: jbyuan@hynu.edu.cn}}
\affiliation{Key Laboratory of Opto-electronic Control and Detection Technology of University of Hunan Province, and College of Physics and Electronic Engineering, Hengyang Normal University, Hengyang 421002, China}
\author{Zhi-Min Tang}
\affiliation{Key Laboratory of Opto-electronic Control and Detection Technology of University of Hunan Province, and College of Physics and Electronic Engineering, Hengyang Normal University, Hengyang 421002, China}
\author{Ya-Ju Song}
\affiliation{Key Laboratory of Opto-electronic Control and Detection Technology of University of Hunan Province, and College of Physics and Electronic Engineering, Hengyang Normal University, Hengyang 421002, China}
\author{Shi-Qing Tang}
\affiliation{Key Laboratory of Opto-electronic Control and Detection Technology of University of Hunan Province, and College of Physics and Electronic Engineering, Hengyang Normal University, Hengyang 421002, China}
\author{Zhao-Hui Peng}
\affiliation{Hunan Provincial Key Laboratory of Intelligent Sensors and Advanced Sensor Materials, and Department of Physics, Hunan University of Science and Technology, Xiangtan 411201, China}
\author{Xin-Wen Wang\footnote{ Email: xwwang@hynu.edu.cn}}
\affiliation{Key Laboratory of Opto-electronic Control and Detection Technology of University of Hunan Province, and College of Physics and Electronic Engineering, Hengyang Normal University, Hengyang 421002, China}
\author{Le-Man Kuang\footnote{ Email: lmkuang@hunnu.edu.cn}}
\affiliation{Key Laboratory of Low-Dimensional Quantum Structures and Quantum Control of Ministry of Education, and Department of Physics, Hunan Normal University, Changsha 410081, China}
\affiliation{Synergetic Innovation Academy for Quantum Science and Technology, Zhengzhou University of Light Industry, Zhengzhou 450002, China}
\date{\today}

\begin{abstract}
We investigate the use of a single generalized dephasing qubit for sensing a quantum reservoir, where the antisymmetry of the coupling between the qubit and its reservoir is broken. Our results indicate that in addition to the decay factor encoding channel, the antisymmetry breaking coupling introduces another phase factor encoding channel. We propose an optimal measurement strategy for the generalized dephasing qubit which enables the practical measurement precision to reach the theoretical ultimate precision quantified by the quantum signal-to-noise ratio (QSNR). As an application example, the generalized dephasing qubit is employed to estimate the $s$-wave scattering length of an atomic Bose-Einstein condensate. It is found that the QSNR contributed by the antisymmetry breaking coupling is at least two orders of magnitude higher than the QSNR contributed by the antisymmetry coupling at the millisecond timescale, and the optimal relative error can achieve a scaling $\propto 1/(\chi t)$  in long-term encoding, where $\chi$ represents the relative driving strength and $t$ is the encoding time. Our work opens a way for supersensitive sensing of quantum reservoirs.
\end{abstract}

\maketitle \narrowtext
\end{CJK*}
\section{\label{Sec:1}Introduction}
Any realistic quantum systems inevitably interact with their surrounding quantum reservoirs, leading to quantum decoherence~\cite{Breuer2007}. It is crucial to assess and characterize the quantum reservoirs for both theoretical researches and practical applications such as quantum coherence protection~\cite{Viola1999,Tan2013} and reservoir engineering~\cite{Zoller1993,Zoller1996}. However, for a complex quantum reservoir with large number of degrees of freedom, it is challenging to precisely estimate various parameters that characterize the quantum reservoir.  An effective way to overcome the challenge is the use of quantum probes~\cite{Recati2005,Cirone2009,Benedetti2014,Grasselli2018,Mehboudi2019,Tamascelli2020,Wu2021,Kirkova2021,Adam2022,Khan2022}. A quantum probe is a small and controllable quantum system prepared in a proper initial state. When the quantum probe interacts with the target quantum reservoir, quantum correlations between them will be generated. These correlations may make the probe sensitive to the reservoir's fluctuations which are induced by small changes in the parameter to be estimated. Therefore, information about the parameter may be extracted by performing an appropriate measurement on the probe. In fact, the precision of this estimation has been extensively studied using tools from the quantum parameter estimation theory ~\cite{Helstrom1976,Holevo1982}.  According to the theory, the ultimate precision of any estimation procedure is limited by the quantum Cram\'{e}r-Rao (QCR) bound, which can be quantified by quantum Fisher information (QFI) or the corresponding dimensionless quantum signal-to-noise ratio (QSNR)~\cite{Wei2013,Jing2019}. A larger QFI (QSNR) indicates a higher potential achievable precision.

A single qubit (a two-level system) is the most straightforward quantum probe for estimating parameters of a quantum reservoir, attracting much attention in recent research~\cite{Chu2020,Wolski2020, Porras2017,Shi2020,Blondeau2021,Ai2021,Zheng2022,Xu2023,Saito2023}. If there is no energy exchange between the qubit and its reservoir, namely, the Hamiltonian of the qubit commutes with the interaction Hamiltonian between the qubit and its reservoir, the dynamical behavior of the qubit can be described by the pure dephasing model~\cite{Breuer2007,Zwerger1987}. The pure dephasing model, which is exactly solvable, has been widely applied to detect various properties of reservoirs such as measuring ultra-low temperatures~\cite{Mitchison2020,Razavian2019,Francesca2020,Candeloro2021,Yuan2023}, probing the cutoff frequency of Ohmic reservoirs~\cite{Benedetti2018,Sehdaran2019,Bahrampour2019,Tan2022,Ather2021}, estimating various coupling strengthes~\cite{Tan2022,Ather2021,Tan2020}, and detecting the non-Markovian properties~\cite{Haikka2011,Haikka2013,Yuan2017}. The interaction Hamiltonian in the dephasing model of a single qubit is usually taken to be~\cite{Breuer2007}
\begin{eqnarray}
\label{intham0}
 \hat{H}_{I}=\hat{\sigma}_{z}\sum_{\mathbf{k}}(g_{\mathbf{k}}\hat{b}^{\dag}_{\mathbf{k}}+g^{*}_{\mathbf{k}}\hat{b}_{\mathbf{k}}),
\end{eqnarray}
where $\hat{\sigma}_{z}=|1\rangle\langle1|-|0\rangle\langle0|$ is the Pauli operator with $|1\rangle$ ($|0\rangle$) being upper (lower) energy level of the qubit probe, $\hat{b}_{\mathbf{k}}(\hat{b}_{\mathbf{k}}^{\dag})$ represents bosonic annihilation (creation) operator for the $k$-th reservoir mode and $g_{\mathbf{k}}$ is the coupling strength. It should be stressed that the coupling form in Eq.~(\ref{intham0}) has antisymmetry, meaning that the qubit in the lower energy level $|0\rangle$ has an opposite coupling strength with each mode of the reservoir as compared to the qubit in the upper energy level $|1\rangle$. We notice that this antisymmetry coupling (AC) only allows the reservoir's parameter information to be encoded into the qubit's decay factor, resulting in a degradation of sensing precision over time during extended encoding~\cite{Cappellaro2017}.

In this paper, we aim to fully utilize the potential of a single dephasing qubit in estimating the parameters of a quantum reservoir. To achieve this, we break the antisymmetry of the coupling in Eq.~(\ref{intham0}) and assume that the qubit in each energy level couples to each mode of the reservoir in arbitrary coupling strength.  During the encoding process, we find that in addition to the decay factor encoding channel, this antisymmetry breaking coupling (ABC) leads to the reservoir's parameters information being encoded into the qubit's phase factor. As a result, the QFI for a certain reservoir's parameter in the generalized dephasing model is composed of two parts:  one part is the AC-contributed QFI, and the other part is the ABC-contributed QFI. This implies that ABC may improve the estimation precision of a single dephasing qubit for estimating reservoir's parameters. Furthermore, we propose a practical measurement scheme that enables the sensitivity of the generalized dephasing qubit to saturate the QCR bound.

To demonstrate the benefits of employing a generalized dephasing qubit for sensing quantum reservoirs, we propose a system comprising an impurity qubit immersed in an atomic Bose-Einstein condensate (BEC) to simulate the generalized dephasing model. We utilize the dephasing qubit to estimate the $s$-wave scattering length of the BEC, which is a crucial parameter in ultracold gases~\cite{Bloch2008}. To independently quantify the sensing precision irrespective of its values, we employ the dimensionless QSNR instead of QFI for consideration. We investigate the dynamical behaviors of the AC-contributed QSNR and the ABC-contributed QSNR separately. Our findings reveal that the ABC-contributed QSNR is at least two orders of magnitude higher than the AC-contributed QSNR at the millisecond timescale. Moreover, the optimal relative error can achieve a scaling $\propto 1/(\chi t)$  in long-term encoding, indicating that ABC allows the relative driving strength (RDS) $\chi$ and the encoding time $t$ to be utilized as resources for enhancing the sensing precision. Consequently, one can conclude that ABC makes it possible to achieve supersensitive sensing of quantum reservoirs.

\section{\label{Sec:2}the generalized dephasing model}
The Hamiltonian of the generalized dephasing model in this paper is given as
\begin{equation}
\label{hami}
 \hat{H}=\frac{\omega_{0}}{2}\hat{\sigma}_{z}+\sum_{{\bf k}}\omega_{{\bf
k}}\hat{b}^{\dag}_{{\bf k}}\hat{b}_{{\bf k}}+\sum_{i=0,1}|i\rangle\langle i|\sum_{\mathbf{k}}(g_{\mathbf{k}i}\hat{b}^{\dag}_{\mathbf{k}}+g^{*}_{\mathbf{k}i}\hat{b}_{\mathbf{k}}),
\end{equation}
where $\omega_{0}$ is level splitting and $\omega_{{\bf k}}$ is the frequency of the $k$-th  reservoir mode. The third term on the right side of the above equation is the interaction Hamiltonian, where $g_{\mathbf{k}0(1)}$ is the coupling strength between the qubit in state $|0\rangle$~$(|1\rangle)$ and the $k$-th reservoir mode. Hereafter we set $\hbar=1$.

Using the relations $|1\rangle\langle 1|=(I+\hat{\sigma}_{z})/2$ and $|0\rangle\langle 0|=(I-\hat{\sigma}_{z})/2$ and omitting the constant term, the interaction Hamiltonian
 \begin{eqnarray}
\label{intham1}
 \hat{H}_{I}=\sum_{i=0,1}|i\rangle\langle i|\sum_{\mathbf{k}}(g_{\mathbf{k}i}\hat{b}^{\dag}_{\mathbf{k}}+g^{*}_{\mathbf{k}i}\hat{b}_{\mathbf{k}})
\end{eqnarray}
is rewritten as
 \begin{eqnarray}
\label{intham2}
 \hat{H}_{I}=\hat{\sigma}_{z}\sum_{\mathbf{k}}(g_{\mathbf{k}}\hat{b}^{\dag}_{\mathbf{k}}+g^{*}_{\mathbf{k}}\hat{b}_{\mathbf{k}})+\sum_{\mathbf{k}}
(\xi_{\mathbf{k}}\hat{b}^{\dag}_{\mathbf{k}}+\xi^{*}_{\mathbf{k}}\hat{b}_{\mathbf{k}}),
\end{eqnarray}
where $g_{\mathbf{k}}$ is coupling strength between the Pauli operator $\hat{\sigma}_{z}$ of the qubit and the $k$-$\mathrm{th}$  reservoir mode and the second term is effective driving term for the harmonic oscillator modes with $\xi_{\mathbf{k}}$  being the driving strength of the $k$-$\mathrm{th}$ reservoir mode. There exist following relationships between $g_{\mathbf{k}}$, $\xi_{\mathbf{k}}$, $g_{\mathbf{k}0}$ and $g_{\mathbf{k}1}$:
\begin{subequations}
\label{gk}
\begin{align}
g_{\mathbf{k}}=\frac{g_{\mathbf{k}1}-g_{\mathbf{k}0}}{2}, \\
\xi_{\mathbf{k}}=\frac{g_{\mathbf{k}1}+g_{\mathbf{k}0}}{2}.
\end{align}
\end{subequations}
It is clearly observed from Eq.~(\ref{gk}b) that setting $g_{\mathbf{k}0}=-g_{\mathbf{k}1}$ for all modes  will cause the interaction Hamiltonian~(\ref{intham2}) to degenerate into the interaction Hamiltonian~(\ref{intham0}). In this paper, the coupling of the qubit in the lower energy level having an opposite coupling strength with each mode of the reservoir as compared to the qubit in the upper energy level, \emph{i.e.} $g_{\mathbf{k}0}=-g_{\mathbf{k}1}$  is referred to as AC. In the generalized depasing model we consider, the antisymmetry is broken ($g_{\mathbf{k}0}\neq-g_{\mathbf{k}1}$), resulting in the extra effective driving term in Eq.~(\ref{intham2}).

Next, we investigate how the quantum state of the qubit evolves over time under the generalized dephasing model. The initial state of the whole system is assumed to be a product state
 \begin{equation}\label{totstate}
\hat{\rho}_{tot}(0)=\hat{\rho}_{s}(0)\otimes\hat{\rho}_{B}(0),
\end{equation}
 where $\hat{\rho}_{s}(0)=|\psi\rangle\langle \psi|$ with $|\psi\rangle=1/\sqrt{2}\left(|0\rangle +|1\rangle\right)$ being a pure state of the probe and $\hat{\rho}_{B}(0)=\prod_{\mathbf k}\left(1-e^{-\beta\omega_{\mathbf k}}\right)e^{-\beta\omega_{\mathbf k}\hat{b}_{{\mathbf k}}^{\dag }\hat{b}_{{\bf k}}}$ is a thermal state of the reservoir, where $\beta=1/k_{B}T$ with $T$ and $k_{B}$ being the temperature and the Boltzmann constant. Then the evolution state of the qubit probe is given as
\begin{equation}\label{state}
\hat{\rho}_{s}(t)=\frac{1}{2}\left(
                \begin{array}{cc}
                 1 & e^{-i\Phi(t)}e^{-\Gamma(t)} \\
                e^{i\Phi(t)}e^{-\Gamma(t)} & 1 \\
                \end{array}
              \right)
,
\end{equation}
where the decay factor $\Gamma(t)$ is
\begin{equation}
\label{gam}
\Gamma(t)=\sum_{{\bf k}}4|g_{\mathbf{k}}|^{2}\frac{(1-\cos \omega _{\mathbf{k}}t)}{\omega
_{\mathbf{k}}^{2}}\coth \left(\frac{\beta\omega_{\mathbf{k}}}{2}\right),
\end{equation}
and the phase factor $\Phi(t)$ has following expression
\begin{equation}
\label{phs}
\Phi(t)=\omega_{0}t-\sum_{\mathbf{k}}4\mathrm{Re}\left[\frac{\xi_{\mathbf{k}}
g^{*}_{\mathbf{k}}}{\omega_{\mathbf{k}}}\right]t+\sum_{{\bf k}}4\mathrm{Im}\left[\frac{\xi_{\mathbf{k}}g^{*}_{\mathbf{k}}}{\omega_{\mathbf{k}}^{2}}
\left(1-e^{-i\omega_{\mathbf{k}}t}\right)\right].
\end{equation}
The derivation of the time-dependent state $\hat{\rho}_{s}(t)$, as given in Eq.~(\ref{state}), is provided in detail in Appendix~\ref{appa}. In the generalized dephasing model, the reservoir's parameter information can be encoded both in the decay factor~(\ref{gam}) and in the phase factor~(\ref{phs}). However, for the dephasing model with AC which is commonly employed in the study of quantum sensing of reservoirs, only the decay factor encodes reservoir parameters due to the absence of an effective driving term ($\xi_{\mathbf{k}}=0$). The ABC in the generalized dephasing model introduces an additional phase factor encoding channel.

\section{\label{Sec:3} optimal measurement for the generalized dephasing model}
In the quantum parameter estimation theory, the estimation precision of a parameter $\lambda$  of interest is restricted to the QCR bound
 \begin{equation}
\delta \lambda \geq \frac{1 }{\sqrt{\nu  \mathcal{F}^{Q}_{\lambda}}},\label{QCR}
\end{equation}
where $\delta \lambda$ is the mean square error, $\nu$ represents the number of repeated experiments and $\mathcal{F}^{Q}_{\lambda}$ denotes QFI with respect to the parameter $\lambda$ . The QFI represents theoretically ultimate precision for single measurement, which can be obtained from the quantum state of the system, and more specifically from its eigenvalues and eigenvectors.  Any qubit state in the Bloch sphere representation can be written as $\hat{\rho}=1/2(\hat{I}+\mathbf{w}\cdot\mathbf{\hat{\sigma}})$, where $\hat{I}$ is $2\times2$ identity matrix, $\mathbf{w}=(w_{x},w_{y},w_{z})^{T}$ is the real Bloch vector and $\mathbf{\hat{\sigma}}=(\hat{\sigma}_{x},\hat{\sigma}_{y},\hat{\sigma}_{z})$ represents the pauli matrices. The eigenvalues of the density operator $\rho$ can be given as $(1\pm w)/2$, where $w$ is the length
of the Bloch vector. The length $w=1$ for pure state and $w<1$ for mixed state. In the Bloch sphere representation the QFI with respect to the estimated parameter $\lambda$ can be given as follows \cite{Wei2013, Jing2019}
\begin{equation}\label{fisher1}
 \mathcal{F}^{Q}_{\lambda}=\left\{
                         \begin{array}{ll}

                            \left|\partial_{\lambda}\mathbf{w}\right|^{2}, & \hbox{$w=1;$} \\
                           \left|\partial_{\lambda}\mathbf{w}\right|^{2}+\frac{\left(\mathbf{w}\hspace{0.05cm}\cdot \hspace{0.05cm}\partial_{\lambda}\mathbf{w}\right)^{2}}{1-w^{2}}, & \hbox{$w < 1$.}

                         \end{array}
                       \right.,
\end{equation}
where $\partial_{\lambda}$ denotes the derivative with respect to the estimated parameter $\lambda$. Therefore, to obtain the QFI of the evolution  state $\hat{\rho}_{S}(t)$ in Eq. (\ref{state}),
the Bloch vector of $\hat{\rho}_{S}(t)$ is given as
\begin{equation}\label{w}
\mathbf{w}=(\cos\Phi e^{-\Gamma},\sin\Phi e^{-\Gamma},0)
\end{equation}
with the length $w= e^{-\Gamma}$. By substituting the Bloch vector in Eq. (\ref{w}) into the Eq. (\ref{fisher1}), the concrete expression of the QFI of the evolution state in Eq. (\ref{state}) is obtained
\begin{equation}\label{fisher2}
 \mathcal{F}^{Q}_{\lambda}=\frac{\left(\partial_{\lambda}\Gamma\right)^{2}}{e^{2\Gamma}-1}+e^{-2\Gamma}\left(\partial_{\lambda}\Phi\right)^{2}.
\end{equation}
The QFI~(\ref{fisher2}) contains of two terms. The first term is the QFI of the pure dephasing model with AC, which  has been extensively studied~\cite{Razavian2019,Francesca2020,Candeloro2021,Yuan2023,Benedetti2018}. Therefore, we can say that the QFI of the second term is contributed by the ABC.
In this paper, the first term is called AC-contributed QFI labeled as
\begin{equation}\label{dfisher2}
 \mathcal{F}^{\parallel}_{\lambda}=\frac{\left(\partial_{\lambda}\Gamma\right)^{2}}{e^{2\Gamma}-1},
\end{equation}
and the second term is called ABC-contributed QFI denoted as
\begin{equation}\label{pfisher2}
 \mathcal{F}^{\perp}_{\lambda}=e^{-2\Gamma}\left(\partial_{\lambda}\Phi\right)^{2}.
\end{equation}

 Now we introduce an optimal measurement scheme for the generalized dephasing model, which enables the sensitivity of the qubit sensor to saturate the QCR bound. For a two-level system, the Fisher information associated with the measurement can be presented as~\cite{Mitchison2020}
\begin{equation}\label{fisher3}
\mathcal{F}_{\lambda}=\frac{1}{\langle\Delta \hat{X}^{2}\rangle}\left(\frac{\partial\langle\hat{X}\rangle}{\partial \lambda}\right)^{2},
\end{equation}
where $\langle\hat{X}\rangle$ and $\langle\Delta \hat{X}^{2}\rangle$ are mean and variance of the measured observable. For an unbiased estimator, the error obeys $\delta \lambda \geq 1/\sqrt{\nu  \mathcal{F}_{\lambda}}\geq 1/\sqrt{\nu  \mathcal{F}^{Q}_{\lambda}}$, which indicates the QFI is the upper bound of the Fisher information associated with the measurement $\hat{X}$, \emph{i.e.}
\begin{equation}
\mathcal{F}^{Q}_{\lambda}=\max_{\hat{X}}\mathcal{F}_{\lambda}(\hat{X})=\mathcal{F}_{\lambda}(\hat{\Lambda})
\end{equation}
with $\hat{\Lambda}$ being the optimal measurement. Finding the optimal measurement $\hat{\Lambda}$ to spur practical precision to reach the theoretically ultimate precision is of particular importance and challenge in quantum metrology. For this reason, we introduce a measurement with an measurement angle $\theta$
\begin{eqnarray}
\label{meas}
\hat{X}_{\theta}=\cos\theta\hat{\sigma}_{x}+\sin\theta\hat{\sigma}_{y},
\end{eqnarray}
where the angle $\theta$ is chosen by the measurer. Then the Fisher information associated with the measurement $\hat{X}_{\theta}$ reads
 \begin{eqnarray}\label{cfisher}
\mathcal{F}_{\lambda}(\hat{X}_{\theta})=\frac{\left[(\partial_{\lambda}\Phi)\sin(\theta-\Phi)-(\partial_{\lambda}\Gamma)\cos(\theta-\Phi)\right]^{2}}
{e^{2\Gamma}-\cos^{2}(\theta-\Phi)}.
\end{eqnarray}
 It is found that when the  angle $\theta$ is chosen to be the phase factor $\Phi$, the Fisher information is equal to the AC-contributed QFI $\mathcal{F}^{\parallel}_{\lambda}$, and when the angle $\theta$ is chosen to be  $\Phi+\pi/2$, the Fisher information is equal to the ABC-contributed QFI $ \mathcal{F}^{\perp}_{\lambda}$, \emph{i.e.}
\begin{equation}
\mathcal{F}_{\lambda}(\hat{X}_{\theta=\Phi})= \mathcal{F}_{\lambda}^{\parallel}, \hspace{0.5cm} \mathcal{F}_{\lambda}(\hat{X}_{\theta=\Phi+\frac{\pi}{2}})=\mathcal{F}_{\lambda}^{\perp}.
\end{equation}
It is further found that when the angle $\theta$ is set to be $\Phi+\varphi$, where $\varphi$ satisfies the following equation
\begin{equation}\label{vphi}
\tan\varphi=\frac{(e^{-2\Gamma}-1)\partial_{\lambda}\Phi}{\partial_{\lambda}\Gamma},
\end{equation}
the Fisher information is equal to the QFI
\begin{equation}\label{eq}
\mathcal{F}_{\lambda}(\hat{X}_{\theta=\Phi+\varphi})=\mathcal{F}^{Q}_{\lambda}.
\end{equation}
See Appendix~\ref{appb} for the detailed verification process. Equation~(\ref{eq}) demonstrates that the measurement $\hat{X}_{\theta=\Phi+\varphi}$ does be the optimal measurement $\Lambda$ which enables the sensitivity of the generalized dephasing qubit sensor to saturate the QCR bound. From Eq.~(\ref{vphi}), we see that the optimal measurement depends on the true value of the parameter. Therefore, measures require some prior information about the parameter $\lambda$, for which a larger sample is required. Meanwhile, the optimal measurement is time dependent, thus precise time control is necessary to achieve the predetermined optimal measurement precision.

\section{\label{Sec:3} quantum sensing to an atomic Bose-Einstein condensate}
\subsection{Quantum simulation of the generalized dephasing model}
 In this subsection we propose a system involving an impurity qubit immersed in a three-dimensional homogeneous atomic BEC to simulate the generalized dephasing model. The qubit probe is confined in a harmonic trap $V_A({\bf r})=m_A\omega_{A}^{2}r^{2}/2$ that is independent of the internal states, where $m_A$ is the mass of the impurity and $\omega_{A}$ is the trap frequency. For $\omega_{A}\gg k_{B}T$, the spatial wave function of the qubit is the ground state of $V_{A}(\mathbf{r})$, \emph{i.e.} $\varphi_{A}({\bf r})=\pi^{-3/4}\ell_{A}^{-3/2}\exp[- r^{2}/(2\ell_{A}^{2})]$
with $\ell_{A}=\sqrt{1/(m_{A}\omega_{A})}$. The Hamiltonian of the qubit is
 \begin{equation}
\label{HA}
 \hat{H}_{A}=\frac{\Omega_{A}}{2}\hat{\sigma}_{z},
\end{equation}
 where $\Omega_{A}$ is level splitting between the lower ($|0\rangle$) and upper ($|1\rangle$) energy levels.
 The Hamiltonian of the BEC is given as
\begin{eqnarray}
\label{HB1} \hat{H}_{B}=&\int d\mathbf{r}\hat{\psi}^{\dag }(
\mathbf{r})\left(-\frac{\hbar^{2}\nabla_{\mathbf{r}}^{2}}{2m_B}+V(\mathbf{r})-\mu
\right) \hat{\psi}(\mathbf{r})\nonumber\\
&+\frac{1}{2}g_{B}\int d\mathbf{r}
\hat{\psi}^{\dag}\left( \mathbf{r}\right) \hat{\psi}^{\dag }\left(
\mathbf{r}\right)\hat{\psi}\left(\mathbf{r}\right)\hat{\psi}\left(\mathbf{r}\right),
\end{eqnarray}
where $\hat{\psi}(\mathbf{r})$ and $\hat{\psi}^{\dag}(\mathbf{r})$ are field creation and annihilation operators, satisfying the bosonic commutative relations
$[\hat{\psi}(\mathbf{r}),\hat{\psi}^{\dag}(\mathbf{r^{'}})]=\delta(\mathbf{r}-\mathbf{r^{'}})$
and
$[\hat{\psi}(\mathbf{r}),\hat{\psi}(\mathbf{r^{'}})]=[\hat{\psi}^{\dag}(\mathbf{r}),\hat{\psi}^{\dag}(\mathbf{r^{'}})]=0$, $\mu$ is chemical potential, the contact interaction strength
\begin{equation}\label{gb}
g_{B}=\frac{4\pi a_{B}}{m_{B}}
\end{equation}
with $a_{B}$ being the $s$-wave scattering length between one condensate atom and another and $m_{B}$ being the mass of the condensate atom. Here we consider $V(\mathbf{r})$ is a very shallow spherical harmonic potential that the condensate can be taken as a homogeneous condensate. Then the field operator can be written as
\begin{eqnarray}
\label{psi}
\hat{\psi}(\mathbf{r})=\sqrt{n}+\delta\hat{\psi}(\mathbf{r}).
\end{eqnarray}
where $n$ is the density of the condensed part and $\delta\hat{\psi}(\mathbf{r})$ is the field operator of small non-condensed part. One can expand $\delta\hat{\psi}(\mathbf{r})$  as plane waves and perform  Bogoliubov transformation
\begin{eqnarray}
\label{dpsi}
\delta\hat{\psi}(\mathbf{r})=\frac{1}{\sqrt{\mathcal{V}}}\sum_{\mathbf{k}}\left(u_{\mathbf{k}}\hat{b}_{{\bf
k}}-v_{\mathbf{k}}\hat{b}^{\dag}_{-{\bf k}}\right)e^{
i\mathbf{k}\cdot\mathbf{r}},
\end{eqnarray}
where $\mathcal{V}$ is the volume of the BEC, $\hat{b}_{\mathbf{k}}$($\hat{b}^{\dag}_{\mathbf{k}}$) is bosonic annihilation (creation) operator satisfying the bosonic commutative relations. Here $u_{\mathbf{k}}$ and $v_{\mathbf{k}}$ are Bogoliubov transformation coefficients with the forms $u_{{\bf k}}=1/2\left(\sqrt{\omega_{\mathbf{k}}/E_{{\bf k}}}+\sqrt{E_{{\bf k}}/\omega_{\mathbf{k}}}\right)$, $v_{{\bf k}}=1/2\left(\sqrt{\omega_{\mathbf{k}}/E_{{\bf k}}}-\sqrt{E_{{\bf k}}/\omega_{\mathbf{k}}}\right)$, where kinetic energy $E_{\mathbf{k}}=k^{2}/(2m_B)$ and Bogoliubov excitation energy $\omega_{\mathbf{k}}$ reads
\begin{eqnarray}
 \omega_{\mathbf{k}}=\sqrt{E_{{\bf k}}^{2}+2ng_{B}E_{{\bf k}}}.
\end{eqnarray}
By substituting Eqs.~(\ref{psi}) and (\ref{dpsi}) into the Hamiltonian (\ref{HB1}) and omitting the constant term and higher-order terms of $\delta\hat{\psi}(\mathbf{r})$, the Hamiltonian of the BEC is diagonalized as~\cite{Oosten2001}
 \begin{equation}
\label{HB}
 \hat{H}_{B}=\sum_{{\bf k}}\omega_{\mathbf{k}}\hat{b}^{\dag}_{{\bf
k}}\hat{b}_{{\bf k}}.
\end{equation}

Let us consider the interaction Hamiltonian between the qubit probe and the BEC. For the qubit-BEC coupling, we assume that the qubit undergoes spin-dependent $s$-wave elastic collisions with the BEC ~\cite{Recati2005,Cirone2009,Haikka2013,Mitchison2020}. The qubit-BEC interaction Hamiltonian is represented as
\begin{equation}\label{Hint1}
\hat{H}_{I}=\left(\sum_{i=0,1}g_{i}|i\rangle\langle i|\right)\int d\mathbf{r}|\varphi_{A}({\bf r})|^{2}\hat{\psi}^{\dag }(
\mathbf{r})\hat{\psi}(\mathbf{r}),
\end{equation}
where
\begin{equation}\label{gi}
g_{i}=\frac{2\pi a_{i}}{m_{AB}}
\end{equation}
is the coupling strength of the qubit-BEC interaction with $a_{i}$ being the $s$-wave scattering length between one condensate atom and the impurity qubit in state $|i\rangle$ and $m_{AB}$ being reduced mass $m_{AB}=m_{A}m_{B}/(m_{A}+m_{B})$. Since the interaction Hamiltonian (\ref{Hint1}) commutes with the Hamiltonian of the qubit in Eq.~(\ref{HA}), the dynamical behavior of the qubit is described by the pure dephasing model. Substituting Eqs.~(\ref{psi}) and (\ref{dpsi}) into the interaction Hamiltonian (\ref{Hint1}) and omitting the constant term and the second order term of $\delta\hat{\psi}(\mathbf{r})$, we obtain
\begin{eqnarray}
\hat{H}_{I}=\frac{\Delta}{2}\sigma_{z}+\sigma_{z}\sum_{\mathbf{k}}g_{\mathbf{k}}\left(\hat{b}_{\mathbf{k}}+\hat{b}^{\dag}_{\mathbf{k}}\right)
+\sum_{\mathbf{k}}\xi_{\mathbf{k}}\left(\hat{b}_{\mathbf{k}}+\hat{b}^{\dag}_{\mathbf{k}}\right),
\end{eqnarray}
where $\Delta=n(g_{1}-g_{2})$, the coupling strength $g_{\mathbf{k}}$ and the driving strength $\xi_{\mathbf{k}}$ are given as
\begin{subequations}
\begin{align}
\label{gx}
g_{\mathbf{k}}=\frac{\sqrt{n}(g_{1}-g_{0})}{\sqrt{\mathcal{V}}}\sqrt{\frac{E_{{\bf k}}}{\omega_{\mathbf{k}}}}e^{\frac{-k^{2}\ell_{A}^{2}}{4}}, \\
\xi_{\mathbf{k}}=\frac{\sqrt{n}(g_{1}+g_{0})}{\sqrt{\mathcal{V}}}\sqrt{\frac{E_{{\bf k}}}{\omega_{\mathbf{k}}}}e^{\frac{-k^{2}\ell_{A}^{2}}{4}}.
\end{align}
\end{subequations}
Let $\Omega_{A}+\Delta=\omega_{0}$, the total Hamiltonian is
 \begin{equation}
\label{Htot}
 \hat{H}=\frac{\omega_{0}}{2}\hat{\sigma}_{z}+\sum_{{\bf k}}\omega_{\mathbf{k}}\hat{b}^{\dag}_{{\bf
k}}\hat{b}_{{\bf k}}+\sigma_{z}\sum_{\mathbf{k}}g_{\mathbf{k}}\left(\hat{b}_{\mathbf{k}}+\hat{b}^{\dag}_{\mathbf{k}}\right)
+\sum_{\mathbf{k}}\xi_{\mathbf{k}}\left(\hat{b}_{\mathbf{k}}+\hat{b}^{\dag}_{\mathbf{k}}\right),
\end{equation}
which confirms that the proposed system successfully simulates the generalized dephasing model. In fact, the generalized dephasing model has been used to study the dephasing dynamics of an impurity qubit in an atomic BEC reservoir~\cite{Cirone2009,Haikka2011,Haikka2013}. It is emphasized that the ratio of the driving strength $\xi_{\mathbf{k}}$ in Eq.~(\ref{gx}b) to the coupling strength $g_{\mathbf{k}}$ in Eq.~(\ref{gx}a) is a physical quantity that is independent of $\mathbf{k}$. We refer to this ratio as the RDS $\chi$ with following form
\begin{eqnarray}\label{chi}
\chi=\frac{\xi_{\mathbf{k}}}{g_{\mathbf{k}}}=\frac{a_{1}+a_{0}}{a_{1}-a_{0}}.
\end{eqnarray}
The above equation shows the RDS $\chi$ can be widely adjusted by changing the $s$-wave scattering lengths $a_{0}$ and $a_{1}$ via Feshbach resonance~\cite{Chin2010}.

\subsection{ Estimating the $s$-wave scattering length $a_{B}$ of the BEC}
In this subsection, we will demonstrate the advantages of the generalized dephasing model in quantum sensing of quantum reservoirs. To illustrate, we employ the impurity qubit to estimate the $s$-wave scattering length $a_{B}$ of the BEC, a crucial parameter in ultracold gases~\cite{Bloch2008}. The initial state of the impurity qubit and BEC is prepared to be the state in Eq.~(\ref{totstate}). Under the control of the Hamiltonian ~(\ref{Htot}), the evolution state of the impurity qubit must be the state in Eq.~(\ref{state}). Substituting $g_{\mathbf{k}}$ in Eq.~(\ref{gx}a) and $\xi_{\mathbf{k}}$ in Eq.~(\ref{gx}b) into the decay factor in Eq.~(\ref{gam}) and the phase factor in Eq.~(\ref{phs}), then  using the continuum limit  $\sum_{{\bf k}}\rightarrow\frac{\mathcal{V}}{(2\pi)^{3}}\int_{0}^{2\pi}\texttt{d}\varphi\int_{0}^{\pi}\sin\theta\texttt{d}\theta\int_{0}^{\infty}k^{2}\texttt{d}k$, we obtain the decay factor
\begin{eqnarray}
\label{gam4}
\Gamma(t)=P\int_{0}^{\infty}k^{2}\frac{E_{{\bf k}}(1-\cos \omega _{\mathbf{k}}t)}{\omega_{\mathbf{k}}^{3}}
e^{\frac{-k^{2}\ell_{A}^{2}}{2}}\texttt{d}k,
\end{eqnarray}
and the phase factor
\begin{eqnarray}\label{phi4}
\Phi(t)=\chi P\int_{0}^{\infty}k^{2}\frac{E_{{\bf k}}(\sin \omega _{\mathbf{k}}t-\omega _{\mathbf{k}}t)}{\omega_{\mathbf{k}}^{3}}e^{\frac{-k^{2}\ell_{A}^{2}}{2}}\texttt{d}k,
\end{eqnarray}
where we consider zero temperature reservoir and the parameter $P=2n(g_{1}-g_{0})^{2}/\pi^{2}$.
\begin{figure}[tbp]
\includegraphics[width=0.99\columnwidth]{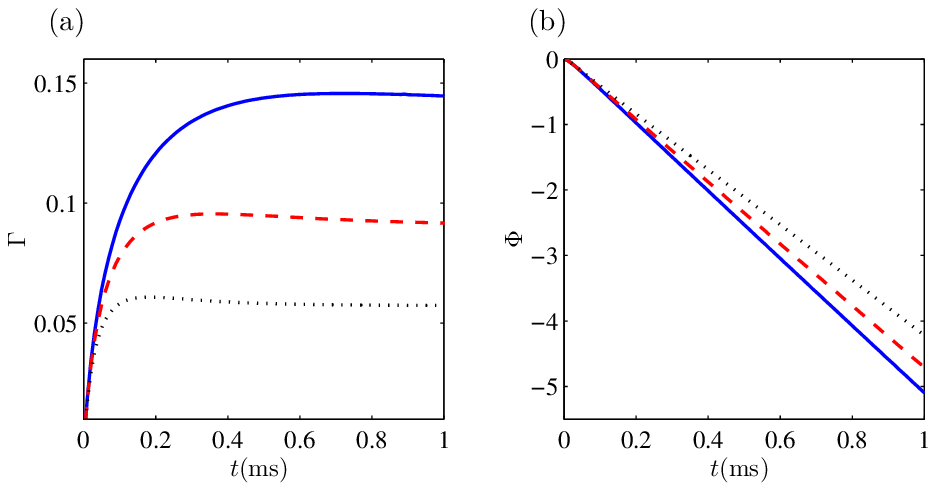}
\caption{(Color online)  Dynamical behaviors of the decay factor $\Gamma$ in (a) and the phase factor $\Phi$ in (b) for the $s$-wave scattering lengths $a_{B}=0.5a_{Rb}$ (blue solid line), $a_{B}=a_{Rb}$ (red dashed line) and  $a_{B}=2a_{Rb}$ (black dotted line). The RDS is taken as $\chi=1$.} \label{fig1}
\end{figure}

We present numerical results based on the Eqs.~(\ref{gam4}) and (\ref{phi4}) after determining reasonable parameter values. We consider a $\mathrm{^{23}Na}$ impurity atom is immersed in a $\mathrm{^{87}Rb}$ BEC with density $n=10^{20}\mathrm{ m^{-3}}$. The impurity atom is trapped in an optical lattice with trapped characteristic length $\ell_{A}=45 \mathrm{nm}$. The difference between the spin-dependent $s$-wave scattering lengths is taken to be $a_{1}-a_{0}=2.9 \mathrm{nm}$. The  $s$-wave scattering length $a_{B}$ of the BEC is restricted by the condition $\sqrt{na_{B}^{3}}<<1$. As a consequence, the scattering length have to satisfy the inequality $a_{B}<3 a_{Rb}$, where $a_{Rb}=5.3 \mathrm{nm}$~\cite{Haikka2011,Song2019}.

The dynamical behaviors of the decay factor $\Gamma$ and the phase factor $\Phi$ for different $s$-wave scattering lengths $a_{B}=0.5a_{Rb}$ (blue solid line), $a_{B}=a_{Rb}$ (red dashed line), and $a_{B}=2a_{Rb}$ (black dotted line) are depicted in Figs.~\ref{fig1}(a) and (b), respectively. The decay factors for the three $s$-wave scattering lengths increase with time from zero and eventually reach different stationary values and the phase factors decease with time, as shown in Figs.~\ref{fig1}(a) and (b). Moreover, the decay factor differences caused by changing $a_{B}$ tend to stabilize over time, but the phase factor differences caused by changing $a_{B}$ become larger with time.  The stationary value $\Gamma(\infty)$ for the  $s$-wave scattering length $a_{B}=0.5a_{Rb}$ is $0.14$, the off-diagonal element $\rho_{10}(t)$ of the density matrix $\hat{\rho}_{s}(t)$ in Eq.~(\ref{state}) will maintain a stable nonzero value $|\rho_{10}(\infty)|=1/2e^{-\Gamma(\infty)}=0.43$. This means that the quantum coherence of the qubit in the BEC reservoir can be preserved at  $|\rho_{10}(\infty)|/|\rho_{10}(0)|=87\%$. We can see that increasing $s$-wave scattering length is beneficial for preserving quantum coherence. The preservation of quantum coherence of a qubit in the BEC reservoir has been addressed in previous studies~\cite{Cirone2009,Haikka2011,Song2019}. Now we introduce the spectral density function to understand the phenomenon of maintaining quantum coherence. The spectral density function is defined as $J(\omega)=\sum_{{\bf k}}4|g_{\mathbf{k}}|^{2}\delta(\omega- \omega _{\mathbf{k}})$. Then the decay factor in Eq.~(\ref{gam4}) can be rewritten as
\begin{eqnarray}\label{gam5}
\Gamma(t)=\int_{0}^{\infty}J(\omega)\frac{1-\cos \omega t}{\omega^{2}}d\omega.
\end{eqnarray}
It is known that the dynamical behavior of the decay
factor $\Gamma(t)$ depends on the specific form of the spectral density function $J(\omega)$~\cite{Breuer2007}.  For example, when the spectral density function is a super-Ohmic spectrum, i.e, $J(\omega)\propto \omega^{s}$ with $s>1$, the decay factor will approach a finite positive value over time, leading to the phenomenon of coherence preservation in dephasing qubits ~\cite{Zwerger1987,Tan2022}. In our proposed system, according to the coupling strength $g_{\mathbf{k}}$ in Eq.~(\ref{gx}), the spectral density function is given as
\begin{eqnarray}\label{spec}
J(\omega)=Q\frac{k(\omega)^{4}}{\omega}\left(\frac{d \omega }{dk}|_{k=k(\omega)}\right)^{-1}e^{-\frac{1}{2}\ell_{A}^{2}k(\omega)^{2}},
\end{eqnarray}
where the parameter $Q=n(g_{1}-g_{2})^{2}/(\pi^{2}m_{B})$ and $k(\omega)$ is obtained from the following dispersion relation
\begin{eqnarray}\label{dr}
\omega\equiv\omega_{\mathbf{k}}=\frac{k\sqrt{k^{2}+16\pi n a_{B}}}{2m_{B}}.
\end{eqnarray}
Different from Ohmic-family spectrum which is phenomenologically given, the specific form of this spectral density in Eq.~(\ref{spec}) depends on the dispersion relation of the BEC in Eq.~(\ref{dr}), and is independent of the ABC. For example, when the wave vector $k<<4\sqrt{\pi n a_{B}}$, the dispersion relation is that of a phonon $\omega= c_{s}k$ with the velocity of sound $c_{s}=2\sqrt{\pi n a_{B}}/m_{B}$. In this case, with $k(\omega)=\omega/c_{s}$, the spectral density function can be approximated as a super-Ohmic spectrum $J(\omega)\propto \omega^{3}$. In short, the structure of the spectral density function, as determined by the dispersion relation of the BEC, leads to the  phenomenon of maintaining quantum coherence.

\begin{figure}[tbp]
\includegraphics[width=0.99\columnwidth]{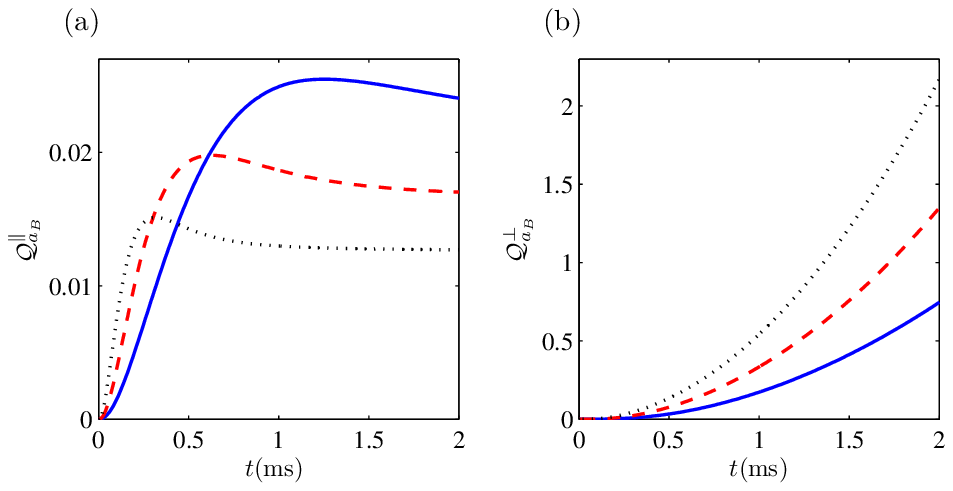}
\caption{(Color online) Time dependence of  the AC-contributed QSNR  $\mathcal{Q}_{a_{B}}^{\parallel}$ in (a) and the ABC-contributed  QSNR $\mathcal{Q}_{a_{B}}^{\perp}$ in (b) for the  $s$-wave scattering lengths $a_{B}=0.5a_{Rb}$ (blue solid line), $a_{B}=a_{Rb}$ (red dashed line) and  $a_{B}=2a_{Rb}$ (black dotted line). The RDS is taken as $\chi=1$.} \label{fig2}
\end{figure}

To quantify the sensing precision of the $s$-wave scattering length $a_{B}$ independently of its values, we introduce the dimensionless QSNR
 \begin{equation}\label{QS}
\mathcal{Q}_{a_{B}}=a_{B}^{2}\mathcal{F}^{Q}_{a_{B}},
\end{equation}
along with the AC-contributed QSNR  $\mathcal{Q}_{a_{B}}^{\parallel}=a_{B}^{2}\mathcal{F}_{a_{B}}^{\parallel}$ and the ABC-contributed QSNR $\mathcal{Q}_{a_{B}}^{\perp}=a_{B}^{2}\mathcal{F}_{a_{B}}^{\perp}$. From the QCR bound in Eq.~(\ref{QCR}), the optimal relative error and the QSNR has the relation
\begin{equation}\label{ore}
\frac{(\delta a_{B})_{min}}{a_{B}}=\frac{1}{\sqrt{\nu\mathcal{Q}_{a_{B}}}},
\end{equation}
which indicates the QSNR quantifies the ultimate precision of quantum sensing. Figures~\ref{fig2}(a) and~\ref{fig2}(b) plot dynamical behaviors of the AC-contributed QSNR $\mathcal{Q}_{a_{B}}^{\parallel}$ and the ABC-contributed QSNR $\mathcal{Q}_{a_{B}}^{\perp}$ for different $s$-wave scattering lengths $a_{B}=0.5a_{Rb}$ (blue solid line), $a_{B}=a_{Rb}$ (red dashed line) and  $a_{B}=2a_{Rb}$ (black dotted line). All AC-contributed QSNRs in Fig.~\ref{fig2}(a) increase from zero to different steady values with time, while all ABC-contributed QSNRs in Fig.~\ref{fig2}(b) increase continuously over time. In particular, it is observed that at the millisecond time scale, the ABC-contributed QSNR  is at least two orders of magnitude higher than the AC-contributed QSNR for the same $s$-wave scattering length $a_{B}$. This demonstrates that the phase factor encoding channel caused by the ABC greatly enhances the ultimate precision of quantum sensing for the $s$-wave scattering length $a_{B}$.  We also need to emphasize that the dynamical behaviors where the QSNR $\mathcal{Q}_{a_{B}}^{\parallel}$ remains unchanged over time and the QSNR $\mathcal{Q}_{a_{B}}^{\perp}$ increases with time both require the preservation of quantum coherence in the qubit.
\begin{figure}
\includegraphics[width=0.65\columnwidth]{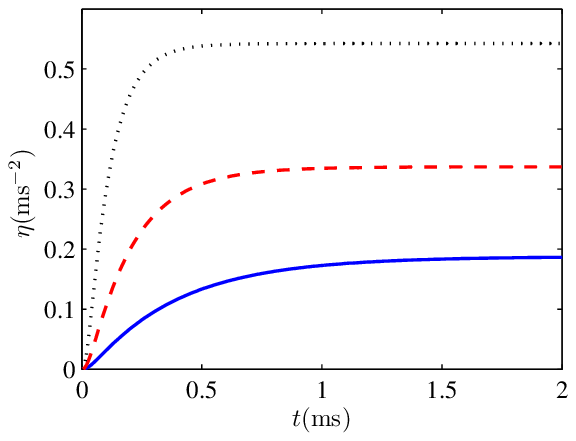}
\caption{(color online) The ratio $\eta$ as a function of  time $t$ for the $s$-wave scattering lengths $a_{B}=0.5a_{Rb}$ (blue solid line), $a_{B}=a_{Rb}$ (red dashed line) and  $a_{B}=2a_{Rb}$ (black dotted line). } \label{ratio}
\end{figure}

\begin{figure}
\includegraphics[width=0.65\columnwidth]{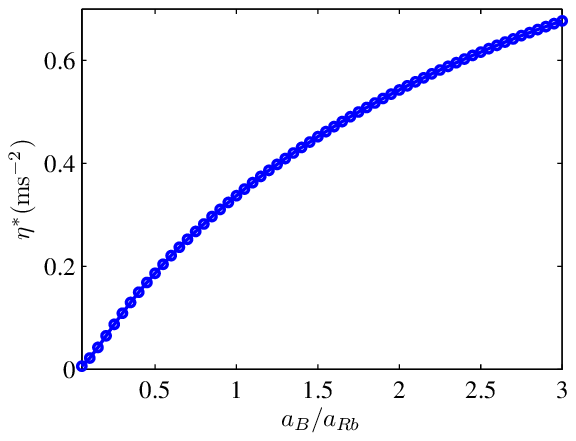}
\caption{(color online) The optimal ratio $\eta^{*}$ as a function of the dimensionless $s$-wave scattering length of the BEC $a_{B}/a_{Rb}$. \label{opratio}}
\end{figure}

To further explore the relationship of the ABC-contributed QSNR $\mathcal{Q}_{a_{B}}^{\perp}$ with the encoding time $t$ and the RDS $\chi$, we define such a ratio
\begin{equation}\label{eta}
\eta=\frac{\mathcal{Q}_{a_{B}}^{\perp}}{(\chi t)^{2}}.
\end{equation}
The dynamical behaviors of the ratio $\eta$ in Eq.~(\ref{eta}) are presented in Fig.~\ref{ratio} for different  $s$-wave scattering lengths $a_{B}=0.5a_{Rb}$ (blue solid line), $a_{B}=a_{Rb}$ (red dashed line) and  $a_{B}=2a_{Rb}$ (black dotted line). As observed, all ratios increase with time to distinct stable values, with the ratio corresponding to a smaller $s$-wave scattering length $a_{B}$ exhibiting a smaller stable value.
In fact, such similar dynamical behaviors are still presented for other $s$-wave scattering lengths. This implies that there exists an optimal ratio $\eta^{*}$ that is independent of the encoding time $t$ and positively correlated with the $s$-wave scattering length $a_{B}$ in long-term encoding. The relationship between the optimal ratio $\eta^{*}$ and the dimensionless $s$-wave scattering length $a_{B}/a_{Rb}$ is depicted in Fig.~\ref{opratio}, which shows that $\eta^{*}$ increases as $a_{B}$ increases.

In long-term encoding, due to $\mathcal{Q}_{a_{B}}^{\perp}\gg\mathcal{Q}_{a_{B}}^{\parallel}$, the QSNR $\mathcal{Q}_{a_{B}}$ can be approximated as
\begin{eqnarray}\label{QA}
\mathcal{Q}_{a_{B}}\approx\mathcal{Q}_{a_{B}}^{\perp}=\eta^{*}(\chi t)^{2},
\end{eqnarray}
thus the optimal relative error has following simple expression
\begin{equation}\label{ore2}
\frac{(\delta a_{B})_{min}}{a_{B}}=\frac{1}{\sqrt{\nu\eta^{*}}\chi t}.
\end{equation}
Equation~(\ref{ore2}) illustrates that encoding time $t$ serves as a resource to augment the ultimate precision of quantum sensing for the $s$-wave scattering length $a_{B}$. Additionally, it demonstrates that increasing the RDS $\chi$ in Eq.~(\ref{chi}) can also enhance the ultimate precision.

\begin{figure}
\includegraphics[width=0.65\columnwidth]{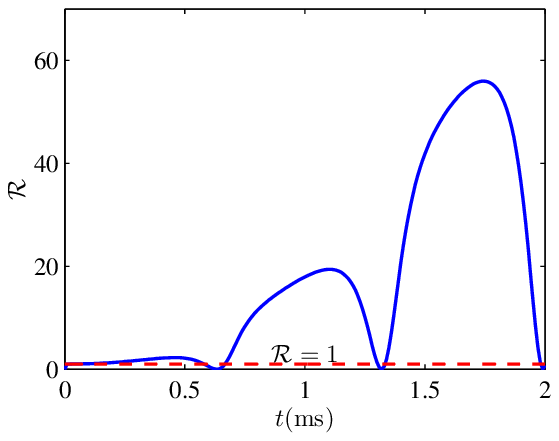}
\caption{(color online) The ratio $\mathcal{R}$ as a function of  time $t$ for the $s$-wave scattering lengths $a_{B}=a_{Rb}$. } \label{ratio2}
\end{figure}

Finally, considering the complexity of the optimal measurement, in order to better demonstrate the superiority of ABC in parameter estimation, we investigate the ratio of the Fisher information associated with the measurement $\hat{\sigma}_{x}$ for the pure dephasing model with ABC to the one for the pure dephasing model with AC
\begin{eqnarray}\label{R}
\mathcal{R}=\frac{\mathcal{F}_{a_{B}}(\hat{\sigma}_{x})}{\mathcal{F}^{\parallel}_{a_{B}}},
\end{eqnarray}
where $\mathcal{F}_{a_{B}}(\hat{\sigma}_{x})$ is given as
\begin{eqnarray}\label{fa}
\mathcal{F}_{a_{B}}(\hat{\sigma}_{x})=\frac{\left[\cos\Phi(\partial_{a_{B}}\Gamma)+\sin\Phi(\partial_{a_{B}}\Phi)\right]^{2}}
{e^{2\Gamma}-\cos^{2}\Phi}.
\end{eqnarray}
Here it should be noted that, for the pure dephasing model with AC, the AC-contributed QFI $\mathcal{F}^{\parallel}_{a_{B}}$ is equivalent to the Fisher information
associated with the measurement $\hat{\sigma}_{x}$~\cite{Razavian2019,Francesca2020,Candeloro2021,Yuan2023,Benedetti2018}. The ratio $\mathcal{R}>1$ means that the ABC can enhance the precision of estimating $a_{B}$ through the measurement $\hat{\sigma}_{x}$, and the larger the value of $\mathcal{R}$, the more significant the effect of enhancing precision. Figure~\ref{ratio2} shows the dynamic behavior of the ratio  $\mathcal{R}$ oscillating over time, which is attributed to the presence of trigonometric functions involving the time-dependent phase factor $\Phi$ in the Fisher information ~(\ref{fa}). Moreover, the local peaks for each oscillation constantly increase with time, as the absolute value of $\partial_{a_{B}}\Phi$ in Eq.~(\ref{fa}) grows with time. Therefore, we can conclude that the ABC can effectively enhance the precision of the s-wave scattering length estimation, attained by the measurement of $\hat{\sigma}_{x}$, for the majority of the time.

\section{\label{Sec:3} Conclusion}
In conclusion, we studied the utilization of a single generalized dephasing qubit for sensing of a quantum reservoir. In the generalized dephasing model, the antisymmetry of coupling between the qubit and its reservoir is broken. Our findings revealed that in addition to the decay factor encoding channel, the ABC introduces another encoding channel, namely the phase factor encoding channel. We employed the QFI to quantify the ultimate precision of quantum sensing and discovered that the QFI associated with the generalized dephasing qubit consists of the AC-contributed QFI and the ABC-contributed QFI. Furthermore, we proposed an optimal measurement scheme for the generalized dephasing qubit, which enables the practical measurement precision to reach the theoretical ultimate precision.

To demonstrate the benefits of employing a generalized dephasing qubit for quantum reservoir sensing, we proposed a system comprising an impurity qubit immersed in an atomic BEC to simulate the generalized dephasing model. We utilized the impurity qubit to estimate the $s$-wave scattering length of the BEC. To independently quantify the sensing precision irrespective of its values, the dimensionless QSNR instead of QFI was employed for analysis. We separately examined the dynamical behaviors of AC-contributed QSNR and ABC-contributed QSNR. Our results indicated that the ABC-contributed QSNR is at least two orders of magnitude higher than the AC-contributed QSNR at the millisecond timescale. Additionally, the ABC-contributed QSNR increases continuously with time, while the AC-contributed QSNR remains constant during long-term encoding. Notably, we discovered that the optimal relative error can achieve a scaling $\propto 1/(\chi t)$ during extended encoding. This means that extending the encoding time $t$ and increasing the RDS $\chi$ can enhance sensing of the $s$-wave scattering length of the BEC. Finally, we have studied the dynamics of the ratio $\mathcal{R}$ , which is defined as  the Fisher information associated with the measurement $\hat{\sigma}_{x}$ under pure dephasing model with ABC to the one under pure dephasing model with AC. It was found that the ABC can also effectively improve the precision of the s-wave scattering length estimation, attained by the measurement of $\hat{\sigma}_{x}$, for the majority of the time.

The generalized dephasing model can also be well used for sensing of other quantum reservoirs such as Ohmic-family reservoirs~\cite{Benedetti2018,Sehdaran2019,Bahrampour2019,Tan2022,Ather2021}. It is worth noting that the ABC enables the encoding of key parameters of the Ohmic-family spectral density, such as the cutoff frequency and reservoir coupling strength, into the phase factor of the generalized dephasing qubit.
Our work opens a way for supersensitive sensing of quantum reservoirs.
\acknowledgments
J. B.  Yuan was supported by  NSFC (No. 11905053) and Scientific Research Fund of Hunan Provincial Education Department of China under Grant (No. 21B0647). L. M. Kuang was supported by NSFC   (Nos. 12247105, 1217050862, and 11935006) and the science and technology innovation Program of Hunan Province (No. 2020RC4047). Y. J. Song was supported by  NSFC (No. 12205088) and Scientific Research Fund of Hunan Provincial Education Department of China under Grant (No. 21B0639). S. Q. Tang was supported by Scientific Research Fund of Hunan Provincial Education Department of China under Grant (No. 22A0507). Z. H.  Peng was supported by  NSFC (No. 11405052) and Hunan Provincial Natural Science Foundation of China under Grant (No. 2020JJ4286).
\begin{widetext}
\appendix
\section{derivation of evolution state of the qubit in the generalized dephasing model \label{appa}}
We first introduce an unitary transformation
\begin{eqnarray}
\hat{U}=\exp\left[\sum_{\mathbf{k}}\left(\hat{\alpha}_{\mathbf{k}}\hat{b}_{\mathbf{k}}^{\dag
}-\hat{\alpha}_{\mathbf{k}}^{\dag}\hat{b}_{\mathbf{k}}\right)\right]
\equiv\Pi_{\mathbf{k}}\hat{U}_{\mathbf{k}}(\hat{\alpha}_{\mathbf{k}}),
\end{eqnarray}
where $\hat{U}_{\mathbf{k}}(\hat{\alpha}_{\mathbf{k}})=\exp\left(\hat{\alpha}_{\mathbf{k}}\hat{b}_{\mathbf{k}}^{\dag
}-\hat{\alpha}_{\mathbf{k}}^{\dag}\hat{b}_{\mathbf{k}}\right)$ is  $k$-th mode unitary transformation
operator with
\begin{eqnarray}
 \hat{\alpha}_{\mathbf{k}}=\frac{
g_{\mathbf{k}}\hat{\sigma}_{z}+\xi_{\mathbf{k}}}{\omega_{\mathbf{k}}}.
\end{eqnarray}
Using the relation
$\exp(\hat{A})\hat{B}\exp(-\hat{A})=\hat{B}+[\hat{A},\hat{B}]+\frac{[\hat{A},[\hat{A},\hat{B}]]}{2!}+\cdot\cdot\cdot$, we obtain
\begin{eqnarray}
\hat{U}_{\mathbf{k}}(\hat{\alpha}_{\mathbf{k}})\hat{b}^{\dag}_{{\bf k}}\hat{b}_{{\bf
k}}\hat{U}^{\dag}_{\mathbf{k}}(\hat{\alpha}_{\mathbf{k}})=\hat{b}^{\dag}_{{\bf k}}\hat{b}_{{\bf
k}}-\hat{\alpha}_{\mathbf{k}}\hat{b}^{\dag}_{\mathbf{k}}-\hat{\alpha}^{\dag}_{\mathbf{k}}\hat{b}_{\mathbf{k}}
+\hat{\alpha}_{\mathbf{k}}\hat{\alpha}^{\dag}_{\mathbf{k}}+\hat{\alpha}^{\dag}_{\mathbf{k}}\hat{\alpha}_{\mathbf{k}},
\end{eqnarray}

\begin{eqnarray}
\hat{U}_{\mathbf{k}}(\hat{\alpha}_{\mathbf{k}})(g_{\mathbf{k}}\hat{b}^{\dag}_{\mathbf{k}}+g^{*}_{\mathbf{k}}\hat{b}_{\mathbf{k}})\hat{U}^{\dag}_{\mathbf{k}}(\hat{\alpha}_{\mathbf{k}})
=(g_{\mathbf{k}}\hat{b}^{\dag}_{\mathbf{k}}+g^{*}_{\mathbf{k}}\hat{b}_{\mathbf{k}})-(g_{\mathbf{k}}\hat{\alpha}^{\dag}_{\mathbf{k}}+g^{*}_{\mathbf{k}}\hat{\alpha}_{\mathbf{k}}).
\end{eqnarray}

Then we perform the unitary transformation $\hat{U}$ on the Hamiltonian (\ref{hami}) of the main text and obtain
\begin{eqnarray}
 \hat{H}^{'}&=&\hat{U}\hat{H}\hat{U}^{\dag}=\frac{1}{2}\omega_{0}\hat{\sigma}_{z}+\sum_{{\bf k}}\omega_{{\bf
k}}\hat{b}^{\dag}_{{\bf k}}\hat{b}_{{\bf k}}\nonumber\\
&&+\sum_{{\bf k}}\left[\omega_{{\bf
k}}(\hat{\alpha}_{\mathbf{k}}\hat{\alpha}^{\dag}_{\mathbf{k}}+\hat{\alpha}^{\dag}_{\mathbf{k}}\hat{\alpha}_{\mathbf{k}})-\hat{\sigma}_{z}(g_{\mathbf{k}}\hat{\alpha}^{\dag}_{\mathbf{k}}+g^{*}_{\mathbf{k}}\hat{\alpha}_{\mathbf{k}})
-(\xi_{\mathbf{k}}\hat{\alpha}^{\dag}_{\mathbf{k}}+\xi^{*}_{\mathbf{k}}\hat{\alpha}_{\mathbf{k}})\right].
\end{eqnarray}

By simplifying the above equation and omitting the constant term, we obtain
\begin{eqnarray}
\label{aham}
 \hat{H}^{'}=\frac{1}{2}\Delta\hat{\sigma}_{z}+\sum_{{\bf k}}\omega_{{\bf
k}}\hat{b}^{\dag}_{{\bf k}}\hat{b}_{{\bf k}},
\end{eqnarray}
where
\begin{eqnarray}
 \Delta=\omega_{0}-\sum_{\mathbf{k}}4\mathrm{Re}\left[\frac{\xi_{\mathbf{k}}g^{*}_{\mathbf{k}}}{\omega_{\mathbf{k}}}\right].
\end{eqnarray}

The evolution state of the qubit probe is represented as
\begin{eqnarray}
 \hat{\rho}_{s}(t)=\mathrm{Tr_{B}}[e^{-i\hat{H}t}\hat{\rho}_{s}(0)\otimes\hat{\rho}_{B}(0)e^{i\hat{H}t}]
 \equiv\mathrm{Tr_{B}}[\hat{U}^{\dag}e^{-i\hat{H}^{'}t}\hat{U}\hat{\rho}_{s}(0)\otimes\hat{\rho}_{B}(0)\hat{U}^{\dag}e^{i\hat{H}^{'}t}\hat{U}].
\end{eqnarray}
According to the above equation, we can easily prove the diagonal elements unchanged. The off-diagonal elements are given as
\begin{eqnarray}
 \rho_{s,10}(t)=\rho_{s,01}(t)^{*}=\langle 1|\hat{\rho}_{s}(t)|0\rangle=\frac{1}{2}e^{-i\Delta t}f_{B}(t).
\end{eqnarray}
Here $f_{B}(t)$ is a reservoir-dependent function with the following expression
\begin{eqnarray}
 f_{B}(t)=\prod_{\mathbf k}\mathrm{Tr_{B}}[\hat{D}_{\mathbf{k}}^{\dag}(\alpha_{\mathbf{k},0})e^{it\omega_{{\bf k}}\hat{b}^{\dag}_{{\bf k}}\hat{b}_{{\bf k}}}\hat{D}_{\mathbf{k}}(\alpha_{\mathbf{k},0})\hat{D}_{\mathbf{k}}^{\dag}(\alpha_{\mathbf{k},1})e^{-it\omega_{{\bf
k}}\hat{b}^{\dag}_{{\bf k}}\hat{b}_{{\bf k}}}\hat{D}_{\mathbf{k}}(\alpha_{\mathbf{k},1})\hat{\rho}_{B,\mathbf{k}}(0)
],
\end{eqnarray}
where $\hat{D}_{\mathbf{k}}(\alpha_{\mathbf{k}})=\exp\left(\alpha_{\mathbf{k}}\hat{b}_{\mathbf{k}}^{\dag}-\alpha_{\mathbf{k}}^{*}\hat{b}_{\mathbf{k}}\right)$ is $k$-th mode Glauber displacement operator with
 $\alpha_{\mathbf{k},1}=(\xi_{\mathbf{k}}+g_{\mathbf{k}})/\omega_{{\bf k}}$, $\alpha_{\mathbf{k},0}=(\xi_{\mathbf{k}}-g_{\mathbf{k}})/\omega_{{\bf k}}$ and $\hat{\rho}_{B,\mathbf{k}}(0)=\left(1-e^{-\beta\omega_{\mathbf k}}\right)e^{-\beta\omega_{\mathbf k}\hat{b}_{{\mathbf k}}^{\dag }\hat{b}_{{\bf k}}}$ is a thermal state of the  $k$-th mode.
Using the following relations
\begin{eqnarray}
\hat{D}(\alpha) \hat{D}(\beta)=\hat{D}(\alpha+\beta)\exp[i\mathrm{Im}(\alpha \beta ^{\ast})],\hspace{0.3cm}
\exp (x\hat{b}^{\dag}\hat{b})\hat{D}(\alpha)\exp(-x\hat{b}^{\dag}\hat{b})=\hat{D}(\alpha
e^{x}),
\end{eqnarray}
we obtain
\begin{eqnarray}
 f_{B}(t)=e^{-i\Theta(t)}e^{-\Gamma(t)},
\end{eqnarray}
where the decaying function is
\begin{eqnarray}
e^{-\Gamma(t)}&=&\prod_{\mathbf k}\mathrm{Tr_{B}}\left[\hat{D}_{\mathbf{k}}\left[(\alpha_{\mathbf{k},1}-\alpha_{\mathbf{k},0})(1-e^{i\omega_{{\bf k}}t})\right]\hat{\rho}_{B,\mathbf{k}}(0)\right]\nonumber\\
&=&\exp\left[-\sum_{{\bf k}}4|g_{\mathbf{k}}|^{2}\frac{(1-\cos \omega _{\mathbf{k}}t)}{\omega
_{\mathbf{k}}^{2}}\coth \left(\frac{\beta\omega_{\mathbf{k}}}{2}\right)\right],
\end{eqnarray}
and the phase factor is given as
\begin{eqnarray}
\Theta(t)&=&\sum_{\mathbf{k}}\mathrm{Im}\left[\alpha_{\mathbf{k},0}(\alpha_{\mathbf{k},0}-\alpha_{\mathbf{k},1})^{*}e^{-i\omega_{{\bf k}}t}+
\alpha_{\mathbf{k},1}^{*}(\alpha_{\mathbf{k},1}-\alpha_{\mathbf{k},0})e^{i\omega_{{\bf k}}t}+2\alpha_{\mathbf{k},0}\alpha_{\mathbf{k},1}^{*}\right]\nonumber\\
&=&\sum_{{\bf k}}4\mathrm{Im}\left[\frac{\xi_{\mathbf{k}}g^{*}_{\mathbf{k}}}{\omega_{\mathbf{k}}^{2}}
\left(1-e^{-i\omega_{\mathbf{k}}t}\right)\right].
\end{eqnarray}

\section{derivation of the fisher information associated with the measurements \label{appb}}
The evolution state in Eq.~(\ref{state}) of the main text is rewritten as
\begin{eqnarray}\label{state2}
\hat{\rho}_{S}(t)=\frac{1}{2}\left(I+e^{-\Gamma(t)}\cos\Phi\hat{\sigma}_{x}+e^{-\Gamma(t)}\sin\Phi\hat{\sigma}_{y}\right).
\end{eqnarray}
We introduce a measurement operator
\begin{eqnarray}
\hat{X}_{\theta}=\cos\theta\hat{\sigma}_{x}+\sin\theta\hat{\sigma}_{y}.
\end{eqnarray}
The mean and the variance of this measurement operator in quantum state~(\ref{state2}) are given as
\begin{eqnarray}
\langle\hat{X}_{\theta}\rangle=e^{-\Gamma}\cos(\theta-\Phi),\hspace{1cm} \langle\Delta\hat{X}_{\theta}^{2}\rangle=1-e^{-2\Gamma}\cos^{2}(\theta-\Phi).
\end{eqnarray}
Here it is important to emphasize that the angle $\theta$ is chosen by the measurer and is not a function of the parameter $\lambda$ to be estimated. Then the Fisher information associated with the measurement $\hat{X}_{\theta}$ reads
 \begin{eqnarray}\label{cfisher}
\mathcal{F}_{\lambda}(\hat{X}_{\theta})=\frac{(\partial_{\lambda}\langle\hat{X}_{\theta}\rangle)^{2}}{\langle\Delta\hat{X}_{\theta}^{2}\rangle}
=\frac{\left[(\partial_{\lambda}\Phi)\sin(\theta-\Phi)-(\partial_{\lambda}\Gamma)\cos(\theta-\Phi)\right]^{2}}
{e^{2\Gamma}-\cos^{2}(\theta-\Phi)}.
\end{eqnarray}

(i) According to the Fisher information in Eq.~(\ref{cfisher}), if one chooses the measurement angle $\theta=\Phi$, the Fisher information equals to AC-contributed QFI, \emph{\emph{i.e.}}

\begin{eqnarray}
\mathcal{F}_{\lambda}(\hat{X}_{\theta=\Phi})=\frac{\left(\partial_{\lambda}\Gamma\right)^{2}}{e^{2\Gamma}-1}=\mathcal{F}^{\parallel}_{\lambda}.
\end{eqnarray}

(ii) If one chooses the measurement angle $\theta=\Phi+\pi/2$,  the Fisher information equals to ABC-contributed QFI, \emph{i.e.}
\begin{eqnarray}
\mathcal{F}_{\lambda}(\hat{X}_{\theta=\Phi+\frac{\pi}{2}})=e^{-2\Gamma}\left(\partial_{\lambda}\Phi\right)^{2}=\mathcal{F}^{\perp}_{\lambda}.
\end{eqnarray}

(iii) We will look for an optimal measurement angle $\theta_{opt}$ that makes the Fisher information $\mathcal{F}_{\lambda}(\hat{X}_{\theta=\theta_{opt}})$ equal to the QFI $\mathcal{F}_{\lambda}^{Q}$.
We assume that the optimal measurement angle can be expressed as $\theta_{opt}=\Phi+\varphi$.  According to the Fisher information in Eq.~(\ref{cfisher}), the Fisher information $\mathcal{F}_{\lambda}(\hat{X}_{\theta=\Phi+\varphi})$ reads
\begin{eqnarray}
\mathcal{F}_{\lambda}(\hat{X}_{\theta=\Phi+\varphi})
=\frac{\sin^{2}\varphi(\partial_{\lambda}\Phi)^{2}
+\cos^{2}\varphi(\partial_{\lambda}\Gamma)^{2}-2\sin\varphi\cos\varphi(\partial_{\lambda}\Phi)(\partial_{\lambda}\Gamma)}
{e^{2\Gamma}-\cos^{2}\varphi}.
\end{eqnarray}
If $\hat{X}_{\theta=\Phi+\varphi}$ is the optimal measurement $\hat{\Lambda}$, it has to satisfy $\mathcal{F}_{\lambda}(\hat{X}_{\theta=\Phi+\frac{\pi}{2}})=\mathcal{F}_{\lambda}^{Q}$, \emph{i.e.}
 \begin{eqnarray}
\frac{\sin^{2}\varphi(\partial_{\lambda}\Phi)^{2}
+\cos^{2}\varphi(\partial_{\lambda}\Gamma)^{2}-2\sin\varphi\cos\varphi(\partial_{\lambda}\Phi)(\partial_{\lambda}\Gamma)}
{e^{2\Gamma}-\cos^{2}\varphi}
=\frac{\left(\partial_{\lambda}\Gamma\right)^{2}}{e^{2\Gamma}-1}+e^{-2\Gamma}\left(\partial_{\lambda}\Phi\right)^{2}.
\end{eqnarray}
By simplifying the above equation, we obtain
\begin{eqnarray}\label{tan}
a\tan^{2}\varphi+b\tan\varphi+c=0,
\end{eqnarray}
where $a=\left(\partial_{\lambda}\Gamma\right)^{2}/(1-e^{-2\Gamma})$, $b=2(\partial_{\lambda}\Phi)(\partial_{\lambda}\Gamma)$ and
$c=(1-e^{-2\Gamma})(\partial_{\lambda}\Phi)^{2}$. By solving equation~(\ref{tan}), we obtain
\begin{eqnarray}
\tan\varphi=\frac{(e^{-2\Gamma}-1)\partial_{\lambda}\Phi}{\partial_{\lambda}\Gamma}.
\end{eqnarray}

\end{widetext}

\end{document}